\documentstyle[12pt,aasms4]{article}
\begin{document}
\newcommand{\msun}{M_{\odot}}
\newcommand{\zsun}{Z_{\odot}}
\newcommand{\kms}{\, {\rm Km\, s}^{-1}}
\newcommand{\cm}{\, {\rm cm}}
\newcommand{\nhi}{N_{HI}}
\newcommand{\lya}{Ly$\alpha$ }
\newcommand{\etal}{et al.\ }
\newcommand{\yr}{\, {\rm yr}}
\title{High Redshift Supernovae and the Metal-Poor Halo Stars:}
\title{Signatures of the First Generation of Galaxies}
\author{Jordi Miralda-Escud\'e$^{1,2,3}$ and Martin J. Rees$^{1,3}$}
\affil{$^1$ Institute for Advanced Study, Princeton, NJ 08540}
\affil{$^2$ University of Pennsylvania, Dept. of Physics and Astronomy,
David Rittenhouse Lab.,
209 S. 33rd St., Philadelphia, PA 19104 (present address)}
\affil{$^3$ Institute of Astronomy, University of Cambridge,
Cambridge CB3 0HA, UK}
\authoremail{jordi@llull.physics.upenn.edu,mjr@ast.cam.ac.uk}

\begin{abstract}

  Recent evidence on the metal content of the high-redshift \lya forest
seen in quasar spectra suggests that an early
generation of galaxies enriched the intergalactic medium (IGM) at
$z\gtrsim 5$. We calculate the number of supernovae that need to have
taken place to produce the observed metallicity. The progenitor stars
of the supernovae should have emitted $\sim 20$ ionizing photons for
each baryon in the universe, i.e., more than enough to ionize the
IGM. We calculate that the rate of these supernovae
is such that about one of them should be observable at any time per square
arc minute. Their fluxes are, of course, extremely faint: at $z=5$,
the peak magnitude should be $K=27$ with a duration of $\sim$ 1 year.
However, these supernovae should still be the brightest objects in the
universe beyond some redshift, because the earliest galaxies should form
before quasars and they should have very low mass, so their luminosities
should be much lower than that of a supernova.

  We also show that, under the assumption of a standard initial mass
function, a significant fraction of the stars in the Galactic halo
should have formed in the early galaxies that reionized and enriched the
IGM, and which later must have merged with our Galaxy. These stars
should have a more extended radial distribution than the observed halo
stars.

\end{abstract}


\section{Introduction}

  Ever since the discovery of the first high-redshift quasar (Schmidt
1965), quasars have maintained their title as the objects with the
highest known redshift; the present record holder is a quasar at
$z=4.89$ (Schmidt, Schneider, \& Gunn 1991). Nevertheless, the highest
known redshifts of galaxies have followed closely behind, with bright
radio galaxies having been found up to $z=4.45$ (Lacy \etal 1994;
Rawlings et al.\ 1996); more
recently, galaxies with high star formation rates have been identified
from interstellar absorption lines at $z=2$ to 3 (Steidel et al.\ 1996)
and from the \lya emission line at $z=4.55$ (Hu \& McMahon 1996).
%Candidate galaxies at $z\sim 6$ were reported from analysis of colors
%(Lanzetta \etal 1996) of objects in the Hubble Deep Field (Williams et
%al.\ 1996), which, if confirmed, would surpass the highest redshift at
%which quasars have been observed.

  In fact, if quasars are related to supermassive black holes that
formed in the centers of high-redshift galaxies, we should expect
that many galaxies already existed before the first quasars appeared.
In any `bottom-up' theory where the observed structure in the universe
forms by hierarchical gravitational collapse, and the primordial
density fluctuations extend to sufficiently small scales, the first
galaxies to form must have had much smaller masses than the present
galaxies. The first stars should have formed in systems with velocity
dispersions of
$\sim 10 \kms$ or lower, corresponding to the lowest temperatures
($T\sim 10^4$ K) that allow cooling and dissipation of the gas
by atomic processes (systems with even lower virial
temperatures can cool and dissipate through molecular hydrogen, but
this cooling process should be suppressed by photodissociation of the
molecules after emission of a number of UV photons that is much smaller
than that needed to reionize the universe; 
see Haiman, Rees, \& Loeb 1996). These systems would be very unlikely
to form quasars, because even a small fraction of their baryons turning
into stars should provide sufficient energy (via ionization, stellar
winds or supernovae) to expel the remaining gas from the shallow
potential well
(e.g., Couchman \& Rees 1986; Dekel \& Silk 1986). Deeper potential
wells, forming at later epochs, are probably needed to form
supermassive black holes in galactic centers.

  Even if all the baryons were converted into stars very efficiently
in these early dwarf galaxies, with a baryonic mass $M_b \lesssim 10^8
\msun$, their total stellar luminosity would be
much smaller than in $L_*$ galaxies at present, simply due to their small
mass. Since only a small fraction of the baryons in these systems is
likely to turn into stars before the gas is ejected, the total stellar mass
in the first galaxies to form in the universe should be much
smaller than $10^8 \msun$.
This implies that a supernova in one of these first galaxies
to form will be far brighter than the galaxy itself. Thus, the
brightest probes of the era when the reionization of the intergalactic
medium (IGM) started
should be supernovae in very small galaxies, caused by the death of the
same stars responsible for the first ionizing photons.
In this paper, we shall estimate the number of supernovae
that should have taken place in these galaxies and should be
observable at very high redshift, and their apparent magnitudes.

\section{Rate of High Redshift Supernovae}

  The scenario where these first small galaxies caused the reionization
of the universe is strongly supported by recent evidence that the metal
abundance in the \lya forest absorption lines with $\nhi \gtrsim 10^{14}
\cm^{-2}$ is typically $Z\sim 10^{-2} \zsun$, from observations of CIV
lines (see Tytler \etal 1995, Songaila \& Cowie 1996 and references
therein). There is relatively little uncertainty in the number of UV
photons that were emitted by the stars that produced a given mass of
heavy elements, because the heavy elements originate from the supernovae
resulting from the same stars that emit most of the ionizing photons
(although the C/O ratio is more uncertain because carbon is more
abundantly produced in lower mass stars). According to the most recent
calculations (Madau \& Shull 1996 and references therein), the energy of
Lyman continuum photons emitted is 0.2\% of the rest-mass energy of the
heavy elements produced. Thus, the energy emitted in ionizing photons
per baryon is $0.002\, m_p c^2 \, \bar Z = 2 \bar Z $ MeV, where $\bar
Z$ is the average metallicity of all baryons in the universe, so only
$\bar Z=10^{-5}$ is needed to have emitted one ionizing photon for each
baryon. If $\bar Z=10^{-2} \zsun = 2\times 10^{-4}$, then 20 ionizing
photons must have been emitted per baryon when the heavy elements were
made. Furthermore, if these were the photons responsible for reionizing
the universe, then each baryon must have recombined 20 times on average
during the reionization epoch. This is a reasonable number, because a
fraction of these photons were probably absorbed in the systems where
the stars were born before the gas was expelled, and those that escaped
could also have been absorbed in Lyman limit systems before the universe
became transparent. Thus, there is no need to invoke ejection of gas
by more massive galaxies that can accrete the ionized IGM to explain
a metallicity $\bar Z = 10^{-2} \zsun$.

  We can now calculate the number of supernovae that were required for
enriching the gas in the IGM to the average metallicity of
$Z=2\times 10^{-4}$ that is observed in the \lya forest at $z\simeq 3$.
This number should depend only on the IGM density and the supernova
yields, and should be independent of any other details related to the
theory for galaxy formation and the type of galaxies that ejected the
enriched gas. Recent simulations of cold dark matter models show that
the absorption lines in the \lya forest can be identified with the
IGM, with density fluctuations caused by gravitational collapse, and
that most of the baryons should be in the IGM in these models
(Cen et al.\ 1994, Hernquist et al.\ 1996, Miralda-Escud\'e et al.\
1996). Thus, it is reasonable to assume that the high-z supernovae
enriched all the baryons in the universe to a mean at metallicity at
least as high as that of the \lya forest.
%This results
%simply from the fact that in these models, and at $z\sim 3$, most of the
%baryons are neither in highly dense, virialized systems, nor in a
%smooth intergalactic medium, but in the filamentary structures forming
%from the collapse of primordial fluctuations having moderate
%overdensities ($1 < \rho/\bar\rho < 100$) which are responsible for the
%\lya forest.
%Notice also that the true metallicity of the IGM may be significantly
%larger than what is inferred from the carbon abundances, because carbon
%is underproduced in Type II supernovae from massive stars relative to
%the solar abundance, and there is evidence for a Si/C ratio in the \lya
%clouds three times larger than solar, as expected for the metals
%produced by a young population (Songaila \& Cowie 1996).

  Since each supernova produces an average of $\sim 1\msun$ of heavy
elements (with uncertainties depending on the assumed initial mass
function and supernova models; see Woosley \& Weaver 1995), this
implies that a supernova
took place at high redshift for each $5000 \msun $ of baryons in the
universe. We shall assume a high baryon density $\Omega_b = 0.025
h^{-2}$, in agreement with the primordial deuterium abundance measured
by Burles \& Tytler (1996) and by Tytler, Fan, \& Burles (1996).
Notice that this implies that most of the
baryons at the present time are dark, so many more baryons than those
we observe in galaxies had to be enriched at high redshift. Assuming
the $\Omega=1$ cosmological model, the total mass of baryons in a
redshift shell of width $\Delta z$ around us is
$M_b = (6c^3\Omega_b)/(GH_0)\,
[1-(1+z)^{-1/2}]^2/(1+z)^{3/2}\, \Delta z$, where $H_0$ is the present
Hubble constant. With the above rate of supernovae per baryon mass
(assumed to take place within the epoch corresponding to the redshift
shell $\Delta z$),
and taking into account that the supernovae within the shell would be
seen by us over a time interval $H_0^{-1} \Delta z / (1+z)^{3/2}$,
the total supernova rate observed over all the sky is 
\begin{equation}
R_{Sup} = 1.8\times 10^8 h^{-2} \, [1-(1+z)^{-1/2}]^2 \yr^{-1} ~.
\end{equation}
or, for $z\sim 5$, about 1 supernova per square arc minute per year. 

\section{Apparent Magnitudes}

  Most of these supernovae should be Type II which, if the progenitor is
a red supergiant, have a plateau of the luminosity in their lightucurves
from 1 to 80 days after the explosion, with $L\simeq 3\times 10^{42}$
erg/s (Woosley \& Weaver 1986). A note of caution should be made here,
in that the low-metallicity progenitors of these early supernovae could
be very different from the high metallicity counterparts. As illustrated
by the case of SN1987A, it is probably not possible at this stage to
predict the type of supernovae we should expect from this first
generation of stars; in particular, the possibility that some supernovae
might reach higher intrinsic luminosities than regular Type IIs should
be kept in mind. A duration of 80 days would be redshifted to more than
a year at redshifts $z>4$, where the supernovae from stars responsible
for the reionization of the universe should occur. At any random time
there should therefore be one or more supernovae per square arc minute
visible in the sky. Of course, many more supernovae should have occurred
in more massive galaxies at later epochs (producing the metals in stars
and interstellar gas at present), and possibly in small systems that
ejected their gas and continued to enrich the IGM.

  The main difficulty in detecting these supernovae is obviously the
extremely faint flux expected. Supernovae should be the brightest
objects in the universe at very high redshift,
but they are of course much fainter than quasars and, as the redshift
increases, the bolometric fluxes decrease at least as rapidly as
$(1+z)^2$. One should point out, however, that when observing at a
fixed, long wavelenth such that the supernovae are observed on the
Rayleigh-Jeans part of the spectrum, the flux actually becomes brighter
as $1+z$, in the limit of high redshift.

  To estimate the apparent magnitudes of the supernovae, we assume a
blackbody spectrum, which is a sufficiently close approximation for our
purpose. In Figure 1 we show the apparent magnitudes in several bands
for a luminosity $L=3\times 10^{42}$ erg/s, and temperatures $T=25000$ K
and $T=7000$ K. The high temperature is reached $\sim$ 1 day after the
explosion, when the luminosity drops to the value in the plateau part of
the lightcurve; during the next few days the temperature cools, reaching
a value near $7000$ K after about a week, and then it stays constant
until the luminosity starts decreasing.
Immediately after the explosion, when the shock reaches the surface of
the star, the luminosity and temperature can be much higher and the
apparent flux can be brighter by a factor $\sim 10$ relative to the
values in Fig. 1, but this phase only lasts for $\sim 1$ hour.
The apparent magnitudes have been calculated for the $\Omega=1$ model
with $H_0 = 70$ Km/s/Mpc, from the flux at the
central wavelengths of the bands, using the central wavelengths and
zero-magnitude fluxes given in Allen (1973). The high-redshift cutoff in
the curves in Figure 1 indicate the redshift where the supernova
light would be absorbed by the \lya forest.

  At high redshift, the supernovae must obviously be searched in the
infrared. The faintest galaxy surveys from the ground have reached
magnitudes $K\simeq 25$ (Cowie et al.\ 1994). From Fig. 1, type II
supernovae would have similar magnitudes at $z\simeq 2.5$, although
searching for variable objects should require more telescope time than
simple object identification. In order to find supernovae at redshifts
higher than known quasars, fainter fluxes by a factor of $5 - 10$ need
to be detected. This might be achieved with implementation of adaptive
optics on large ground-based telescopes; in the longer term, the New
Generation Space Telescope should certainly be capable to observe these
supernovae (see Mather \& Stockman 1996).
We emphasize again the large uncertainty in predicting
the types and absolute magnitude of supernovae from these early
generation of stars.
%Some type I supernovae (which are typically 1 magnitude
%brighter) may already be taking place at that epoch; their rate is of
%course very uncertain, depending on the number of white dwarfs in close
%binary systems that can form early.

  One possibility to detect these supernovae before more powerful
telescopes and cameras in the infrared can be built is to use the
magnifying power
of gravitational lensing in galaxy clusters. The deflection angles
of the most massive clusters of galaxies are as large as $b\sim 30''$,
with critical lines of total length of several arc minutes.
The cross section for magnifying a source by more than a factor $A$
is $\sim \pi b^2/A^2$, or 0.01 square arc minutes for $A=10$. Thus,
any rich lensing cluster should have a 1\% chance of having a
high-redshift supernova magnified by a factor larger than 10 at any
time. The highly magnified images would always appear in pairs around
critical lines, and would be simple to identify only from their
positions and colors given a lensing model of the cluster (see
Miralda-Escud\'e \& Fort 1993). Later, variability would have to be
detected to distinguish the supernova from a faint, compact galaxy.

\section{The Present Distribution of Population III Stars}

  We have suggested in this paper that, under reasonable assumptions,
supernovae should be the brightest objects in the universe
beyond some redshift, in particular during the early phases of the
reionization. The supernovae might therefore be the first observational
evidence we shall have of this epoch, when the very faint apparent
magnitudes expected are observable. The other observable signature
of this epoch may be the 21cm absorption or emission
by the neutral intergalactic gas (Scott \& Rees 1994; Madau, Meiksin \&
Rees 1996).

  The UV and heavy elements abundance inferred from quasar absorption
lines allow us, as we have seen, to draw quantitative conclusions about
the minimum number of high-mass stars formed beyond $z=5$: to produce
a metallicity $Z=2\times 10^{-4}$ requires one supernova for each
$\sim 5000 \msun$ of baryons. This inference is quite robust, being
insensitive to the details of structure and galaxy formation at high
redshift, which of course depend on the cosmological assumptions.
However, the total mass of stars formed depends on the IMF, and is
therefore much more uncertain.
For a standard IMF, $\sim 100 \msun$ of stars need to be formed to
produce one supernova, so 2\% of the baryons should have been turned
into stars by the time the IGM reached this level of enrichment.
Notice that this is equivalent to $20\%$ of the {\it observed}
baryons in galaxies today, given our adopted value of $\Omega_b$ which
implies that only $\sim$ 10\% of the baryons are in known stars and gas
in galaxies. It is of course quite possible that the IMF was different
for these early stars, given the different physical environment (a
higher ambient temperature, absence of heavy elements to act as coolants
and provide opacity, and no significant magnetic fields). Direct clues
to the slope of the high-mass IMF may come from (for instance) the
relative ionization levels of H and He and heavy elements, which depend
on the background radiation spectrum shortward of the Lyman limit, or
from relative abundances of heavy elements relative to carbon.
Conceivably, all the early stars might be of high-mass, so that no
coeval low-mass stars survive; at the other extreme, the early IMF could
have been much steeper than the standard one, in which case there could
be many pregalactic brown dwarfs.

  Would any of these ``Population III'' stars be observable today? Let
us consider the observational consequences of the simplest assumption:
that the early IMF was the same as in the solar neighborhood. In that
case, most of the present luminosity from the Population III
would arise from red giants and stars at the tip of the
main-sequence, with $M\sim 0.8 \msun$. Where should these stars be
today?

  After the first galaxies ejected all their gas back to the IGM, the
stars that had been formed should have remained in orbit near the center
of the dark matter halos. The stars then behave as collisionless matter
as the halos merge with larger objects, until the present galaxies are
formed.
%  There is also the question of the spatial distribution that the
%Population III stars should have. If the baryons in the first galaxies
%had condensed into the center of the potential well following radiative
%dissipation, so that they became self-gravitating (rather than being
%bound mainly by the dark matter), and a brief burst of star formation
%could subsequently have ejected most of the baryons (still in gaseous
%form) in much less than a dynamical time, then the stars could be left
%unbound and escape their potential wells, and would subsequently cluster
%non-dissipatively. At present,
%they would be distributed like the dark matter in halos,
%including the halo of our Galaxy.
We would therefore expect that these stars would at present be
distributed approximately like the dark matter in galactic and cluster
halos, and in addition there should be some surviving galaxies from
that epoch which have not merged into much larger objects (or have
survived in orbit after merging with a large halo, having escaped
tidal disruption) and still
have the Population III stars in their centers.
The halos of stars formed in this way around galaxies might be somewhat
more centrally concentrated than the dark matter, if many mergers take
place with only a moderate increase of the halo mass at each merger
(so that dynamical friction is effective after each merger and it can
bring the stars near the center of the newly formed halo before tidal
disruption occurs). In fact, particles that start near the centers of
halos that merge tend to end up near the center of the merger product
(e.g., Spergel \& Hernquist 1992).
%If a large number of
%successive mergers took place, with only a modest fractional increase
%of the mass of the new halo at each merger, then the stars could have
%time to sink via dynamical friction towards the center of every new
%merged system, preserving to some degree their central concentration
%relative to the dark matter. Alternatively, if the original low-mass
%galaxy remained isolated initially and merged with our galaxy
%at a late time, dynamical friction would be negligible and the merging
%system would either remain in orbit in our galactic halo or be tidally
%disrupted, depending on its density and the pericenter of its orbit.
%The resulting stars or extreme dwarf galaxies should be distributed
%like the dark matter.

  The known halo stars have a very steep density profile,
$\rho \propto r^{-3.5}$, and their total mass is $M\sim 10^9 \msun$
(e.g., Morrison 1993). This
mass is comparable to the total mass we would expect in the halo in the
Population III stars, if the total mass of the halo of our Galaxy is
$5\times 10^{11} \msun$, with a baryon fraction of 10\%, and if 2\% of
the baryons formed Population III stars. Therefore, if the
stellar mass function in the first galaxies was normal, a sizable
fraction of the halo stars should have originated there (this is not
surprising, because it is derived from the assumption that the halo
stars created their own metal abundance). It seems difficult that the
process of dynamical friction alluded to above can result in the steep
slope of the halo stars. However, the halo density profile might become
shallower at large radius (see Hawkins 1983 and Norris \& Hawkins 1991
for current observational evidence on this possibility),
and a second halo population in the
outer part of the galaxy ($R\sim 100$ Kpc) might be the remnant
of the Population III. These halo stars could be found in the Hubble
Deep Field (HDF, Williams \etal 1996). If the stellar mass of this outer
halo is $10^9 \msun$, there should be $\sim 10^8$ stars near the
main-sequence turnoff, i.e., we expect a few stars in the HDF (with area
4.4 arcmin$^2$); these would have colors $I-V \simeq 1.5$, $I\sim 25$ at
distances of 100 Kpc. From Fig. 2 in Flynn, Gould, \& Bahcall, we see that
there is at least one stellar object with these characteristics in the
HDF.
  
  Several other observations may help to test the existence of the
Population III stars. An outer stellar halo would also imply a certain
number of high-velocity stars near the solar neighborhood. A stellar
population may be found in the halos of external galaxies, with density
profiles similar to the dark matter. Sackett \etal (1994)
found a luminous halo in the galaxy NGC 5907 with $M/L = 500$ (i.e.,
about ten times more light than what we expect for the Population III).
Planetary nebulae could also be found in nearby halos of galaxies or
galaxy groups; several of them were reported recently by by Theuns \&
Warren 1996) in the Fornax cluster.

  There is also the possibility that the IMF in the early galaxies
produced a large number of brown dwarfs. In this case, a large fraction
of the baryons could have been turned into brown dwarfs, and these could
be detected in ongoing microlensing experiments towards the LMC
(see Paczy\'nski 1996). If the
baryon fraction in the universe is 10\%, the optical depth of
these brown dwarfs toward the LMC could be as high
as a few times $10^{-8}$.

  Finally, we notice that the metallicity distribution of the
Population III stars is difficult to predict. If only a small fraction
of the neutral IGM collapsed to galaxies before the reionization, then
the gas in these galaxies could reached high metallicities and formed
stars, and the metallicity could have diluted in the IGM when the gas
was ejected. At the same time, the metal abundance of the IGM after
reionization could be highly inhomogeneous, so some galaxies formed
later could have very low metallicities. Therefore, it is difficult
to predict even if the average metal abundance of the Population III
stars should be higher, lower or similar to the more centrally
concentrated halo stars, let alone the distribution of these
metallicities.

%the IGM. We know from the ubiquitous metal lines in the \lya forest
%that the metals must have been sufficiently
%dispersed such that their covering factor through a \lya cloud is
%large. This still allows for a substantial inhomogeneity in
%three-dimensional space. Nevertheless, before this gas may turn into
%stars in a galaxy, it needs to virialize and be shocked and increase
%its density by several orders of magnitude, a process that is likely
%to lead to efficient mixing. In support of this point of view, one may
%notice that the metallicities of globular clusters do not extend to
%values as low as among halo stars (Laird \etal 1988; Ryan \& Norris
%1991).

\section{Conclusions}

  As the observational techniques improve our ability to detect extremely
faint sources, and higher redshift objects can be searched for to
continue unravelling the history of galaxy formation,
supernovae should become the brightest observable sources. These
supernovae created the heavy elements that were expelled to the
IGM, and their progenitor stars are the most likely
sources of the photons that reionized the universe. The expected rates
of these supernovae, calculating under the assumption of a high baryon
density ($\Omega_b h^2 = 0.025$), and an average metal production of
$\bar Z = 10^{-2} \zsun$, is as high as 1 supernova per square arc minute
per year. To detect the supernovae, the flux limits of the faintest
sources detectable with our telescopes will probably need to be pushed
by another $\sim 2$ magnitudes, although the first examples might be
discovered at brighter fluxes behind clusters of galaxies, using the
lensing magnification.

  Any low-mass stars that were formed in the first small galaxies where
these supernovae took place should be observable today. We have argued
that, if the IMF in these galaxies was similar to the present one in
our galactic disk, the Population III stars are likely to account for a
large fraction of the stars in our galactic halo, although most of them
should be in an as yet undetected outer halo with a shallower density
profile than the known, inner stellar halo.

\acknowledgements
We thank Len Cowie, Andy Gould and John Norris for stimulating discussions.
JM acknowledges support by the W. M. Keck Foundation at IAS.

\newpage

\begin{figure}
\centerline{
\hbox{
\epsfxsize=2.4truein
\epsfbox[75 32 525 756]{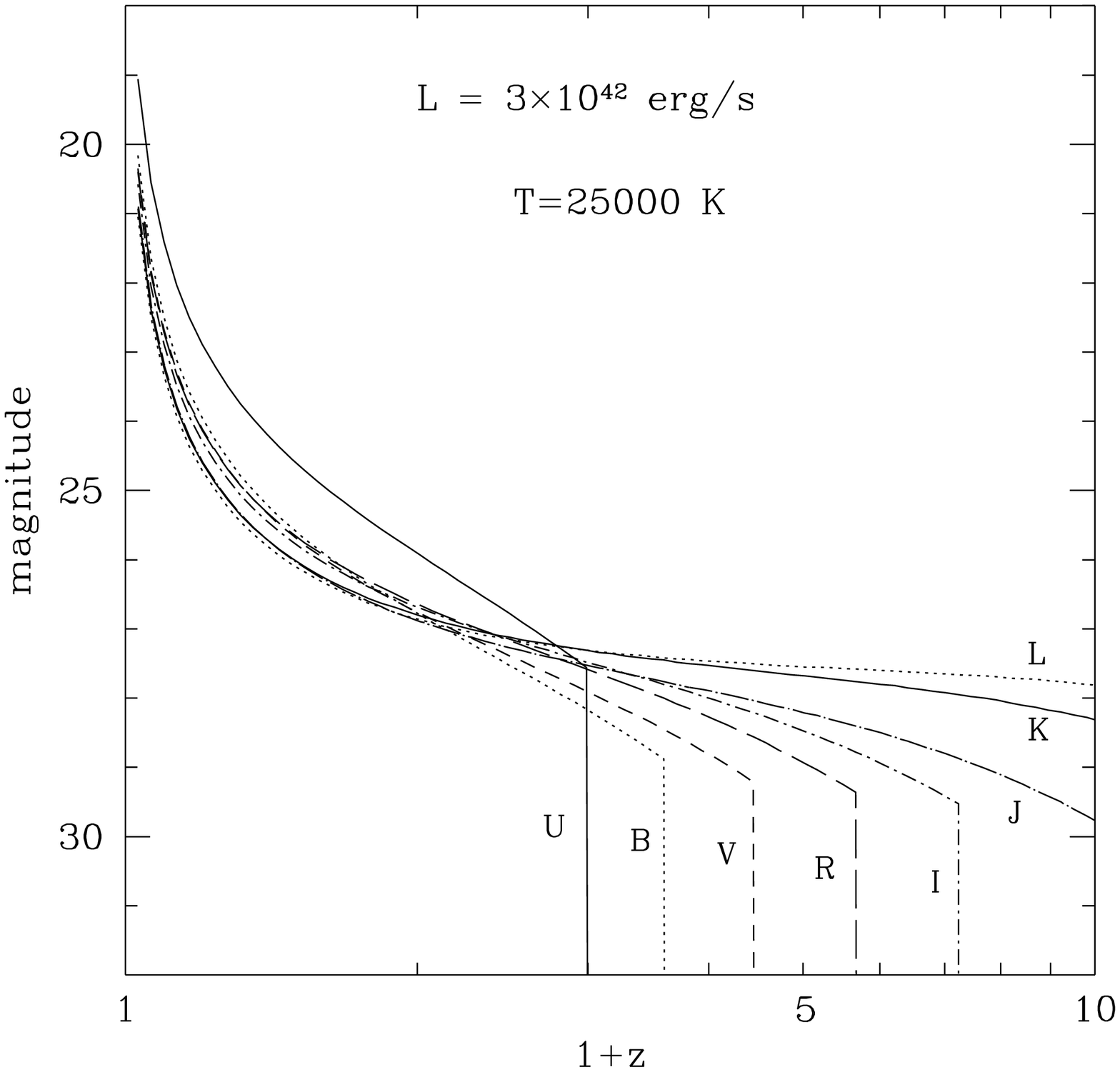}
\hskip 0.75truein
\epsfxsize=2.4truein
\epsfbox[75 32 525 756]{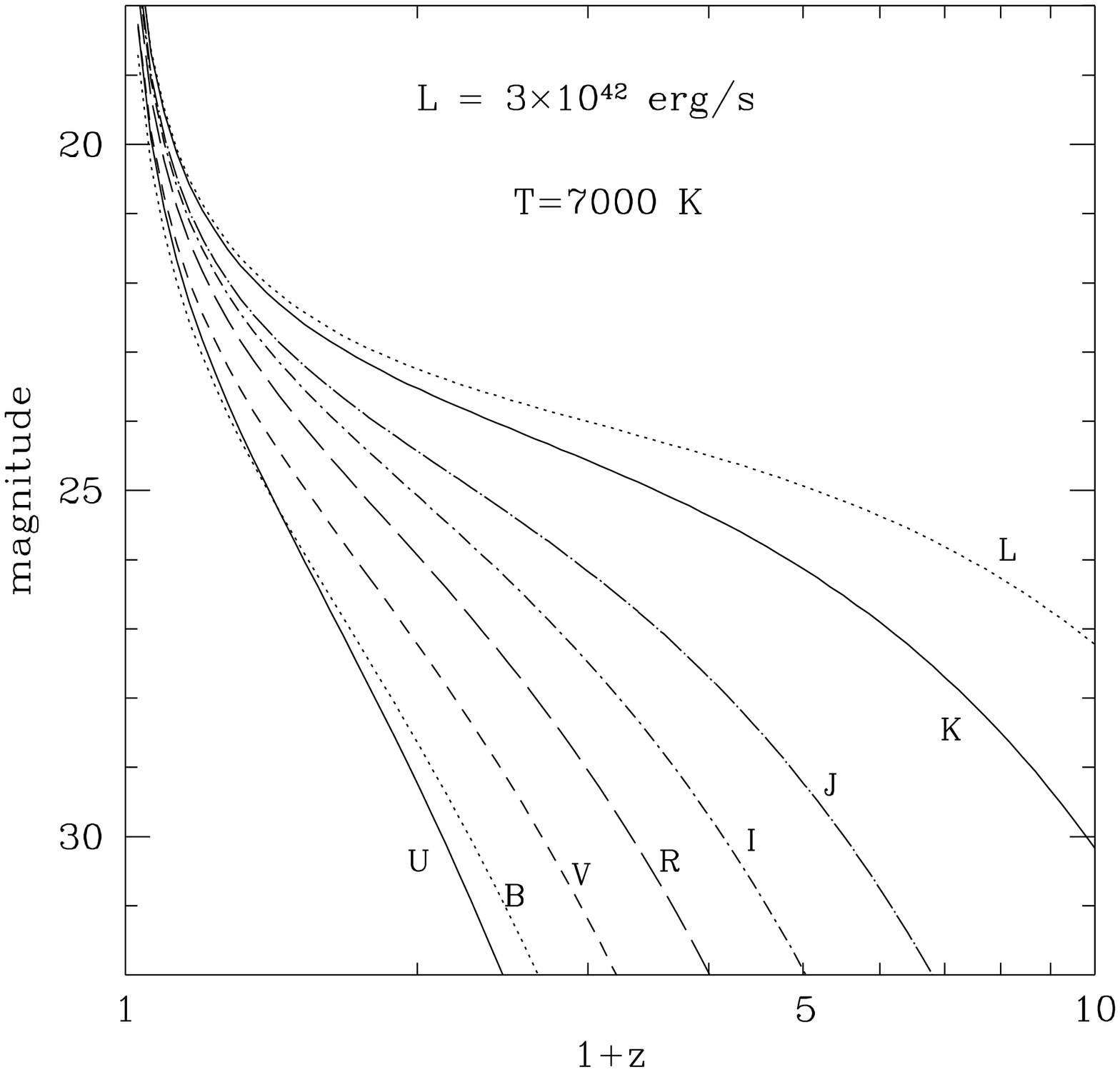}
}
}
\caption{
Apparent magnitude of a supernova as a function of redshift in
different bands, as labeled in the figure, for two different
temperatures. We assume the supernova spectrum
is a blackbody with the luminosity and temperature indicated.
Supernovae Type II generally have a plateau of constant luminosity in
their lightcurves with the value assumed in this figure lasting for
80 days. A temperature of 25000 K is reached $\sim$ two days after
the explosion, and a week later the temperature has dropped to near
7000 K, where it stays constant for the next two months.
}
%%\psfig{file=fig1b.ps,width=17.cm,angle=0.}
%\caption{
%Apparent magnitude of a supernova as a function of redshift in
%different bands, as labeled in the figure, for two different
%temperatures. We assume the supernova spectrum
%is a blackbody with the luminosity and temperature indicated.
%Supernovae Type II generally have a plateau of constant luminosity in
%their lightcurves with the value assumed in this figure lasting for
%80 days. A temperature of 25000 K is reached $\sim$ two days after
%the explosion, and a week later the temperature has dropped to near
%7000 K, where it stays constant for the next two months.
%}
\end{figure}
\vfill\eject

%\end{document}

%\begin{figure}
%\centerline{
%\psfig{file=fig1a.ps,width=17.cm,angle=0.}
%}
%\end{figure}
%\vfill\eject
%
%\begin{figure}[p]
%\centerline{
%\psfig{file=fig1b.ps,width=17.cm,angle=0.}
%}
%\caption{
%Apparent magnitude of a supernova as a function of redshift in
%different bands, as labeled in the figure, for two different
%temperatures. We assume the supernova spectrum
%is a blackbody with the luminosity and temperature indicated.
%Supernovae Type II generally have a plateau of constant luminosity in
%their lightcurves with the value assumed in this figure lasting for
%80 days. A temperature of 25000 K is reached $\sim$ two days after
%the explosion, and a week later the temperature has dropped to near
%7000 K, where it stays constant for the next two months.
%}
%\end{figure}
%\vfill\eject


\begin{thebibliography}{}

\bibitem[]{}Allen, C. W. 1973, {\it Astrophysical Quantities} (London: Athlone Press)
\bibitem[]{} Burles, S., \& Tytler, S. 1996, submitted to Science (astroph 9603069)
\bibitem[]{} Cen, R., Miralda-Escud\'e, J., Ostriker, J. P., \& Rauch, M. 1994,
ApJ, 437, L9
\bibitem[]{} Couchman, H. M. P., \& Rees, M. J., 1986, MNRAS, 221, 53
\bibitem[]{} Cowie, L. L., Gardner, J. P., Hu, E. M., Songaila, A., Hodapp, K.-W., \&
\bibitem[]{} Wainscoat, R. J. 1994, ApJ, 434, 114
\bibitem[]{} Dekel, A., \& Silk, J. 1986, ApJ, 303, 39
\bibitem[]{} Flynn, C., Gould, A., \& Bahcall, J. N. 1996, ApJ, 466, L55
\bibitem[]{} Haiman, Z., Rees, M. J., \& Loeb, A. 1996, ApJ, submitted
(astroph-9608130)
\bibitem[]{} Hawkins, M. S. 1983, MNRAS, 206, 433
\bibitem[]{} Hernquist, L., Katz, N., Weinberg, D. H.,
\& Miralda-Escud\'e, J. 1996, ApJ, 457, L51
\bibitem[]{} Hu, E. M., \& McMahon, R. G. 1996, Nature, 382, 231
\bibitem[]{} Lacy, M., et al.\ 1994, MNRAS, 271, 504
%\bibitem[]{} Laird, J. B., Rupen, M. P., Carney, B. W., \& Latham, D. W. 1988,
%AJ, 96, 1908
%Lanzetta, K. , \& Yahil, A. 1996, preprint
\bibitem[]{} Madau, P., Meiksin, A., \& Rees, M. J. 1996, ApJ, submitted
(astroph 9608010)
\bibitem[]{} Madau, P., \& Shull, J. M. 1996, \apj, 457, 551
\bibitem[]{} Mather, J., \& Stockman, H. 1996, NASA Report.
\bibitem[]{} Miralda-Escud\'e, J., \& Fort, B. 1993, ApJ, 417, 5
\bibitem[]{} Miralda-Escud\'e, J., Cen, R., Ostriker, J. P., \& Rauch, M. 1996, ApJ, 471, 582
\bibitem[]{} Morrison, H. L. 1993, AJ, 106, 578
\bibitem[]{} Norris, J., \& Hawkins, M. S. 1991, ApJ, 380, 104
%Norris, J., \& Ryan, S. 1991 ApJ, 380, 104
%Norris, J. 1994, ApJ, 431, 645
\bibitem[]{} Paczy\'nski, B. 1996, ARA\&A, 34, XXX
\bibitem[]{} Rawlings, S., Lacy, M., Blundell, K. M., Eales, S. A.,
Bunker, A. J., \& Garrington, S. T. 1996, Nature, 383, 502
%\bibitem[]{} Ryan, S. G., \& Norris, J. E. 1991, AJ, 101, 1865
\bibitem[]{} Sackett, P. D., Morrison, H. L., Harding, P., \& Boroson, T. A.
1994, Nature, 370, 441
\bibitem[]{} Schmidt, M. 1965, ApJ, 141, 1295
\bibitem[]{} Schneider, D. P., Schmidt, M., \& Gunn, J. E. 1991, AJ, 101, 2004
\bibitem[]{} Scott, D., \& Rees, M. J. 1990, MNRAS, 247, 510
\bibitem[]{} Songaila, A., \& Cowie, L. L. 1996, AJ, in press (astro-ph 9605102)
\bibitem[]{} Spergel, D. N., \& Hernquist, L. 1992, ApJ, 397, L75
\bibitem[]{} Steidel, C. C., Giavalisco, M., Pettini, M., Dickinson, M., \&
Adelberger, K. L. 1996, ApJ, 462, L17
\bibitem[]{} Theuns, T., \& Warren, S. J. 1996, submitted to MNRAS
(astro-ph 9609076)
\bibitem[]{} Tytler, D., Fan, X.-M., Burles, S., Cottrell, L., Davis, C., Kirkman, D.,
\& Zuo, L. 1995, in QSO Absorption Lines, ed. G. Meylan, p. 289
\bibitem[]{} Tytler, D., Fan, X.-M., \& Burles, S. 1996, Nature, 381, 207
\bibitem[]{} Williams, R., et al. 1996, Science with the Hubble Space
Telescope II, eds. P. Benvenuti, F. D. Macchetto, \& E. J. Schreier
(Baltimore: STScI), in press
\bibitem[]{} Woosley, S. E., \& Weaver, T. A. 1995, ApJS, 101, 181
\bibitem[]{} Woosley, S. E., \& Weaver, T. A. 1986, ARA\& A, 24, 205

\end{thebibliography}
\end{document}